\documentclass[usenatbib]{mn2e}
\usepackage[dvips]{graphicx,color}
\usepackage{amssymb}
\title[IGR J17098-3628 \& EXO 0748-676]
  {Discovery of long-term superorbital periodicities in the pseudo-transient LMXBs: \\
IGR J17098-3628 and EXO 0748-676}
\author[M.M. Kotze et al.]
  {M.M.~Kotze,$^{1,2}$
  P.A.~Charles$^1$, L.A.~Crause$^1$\\
  $^1$ South African Astronomical Observatory,
       P.O. Box 9, Observatory 7935, South Africa\\
  $^2$ Dept. of Maths \& Applied Maths, University of Cape Town, South Africa\\
}
\date{Released 2009 February 10}

\pagerange{\pageref{firstpage}--\pageref{lastpage}} \pubyear{2008}

\def\LaTeX{L\kern-.36em\raise.3ex\hbox{a}\kern-.15em
    T\kern-.1667em\lower.7ex\hbox{E}\kern-.125emX}

\begin{document}

\label{firstpage}

\maketitle

\begin{abstract}
\noindent Long-term monitoring of the recently discovered X-ray transient, IGR J17098-3628, by the All Sky Monitor on board the Rossi X-ray Timing Explorer, has shown that it displays a long term ($\approx$ 163d) quasi-periodic modulation in the data spanning its ``active'' state (i.e. approximately MJD 53450-54200).  Furthermore, this light-curve is not typical of ``classical'' soft X-ray transients, in that J17098-3628 has remained active since its initial discovery, and may be more akin to the pseudo-transient EXO0748-676, which is now classified as a persistent Low Mass X-ray Binary.  However, EXO0748-676 recently entered a more active phase (since approximately MJD 53050), since when we find that it too displays a quasi-periodic modulation ($\approx$ 181d) in its light-curve.  This must be a ``superorbital'' modulation, as the orbital period of EXO0748-676 is well established (3.8hrs), and hence we interpret both objects' long periods as representing some intrinsic properties of the accretion disc (such as coupled precessional and warping effects).  By analogy, we therefore suggest that IGR J17098-3628 is another member of this class of pseudo-transient LMXBs and is likely to have a $<$1d orbital period.

\end{abstract}

\section{Introduction} 

X-ray transients are X-ray sources that flare up suddenly and fade away again after a few weeks, displaying a typical outburst profile in the lightcurve. A steep increase toward maximum flux is followed by a gradual, often exponential decay back toward quiescence (usually $\geq$1000 times fainter for the soft X-ray transients (SXTs), see McClintock \& Remillard 2006, and references therein). Sudden changes in the accretion rate through the accretion disc onto a neutron star or black hole are believed to lead to this behaviour (e.g. King 2006, and references therein). 

X-ray transients are generally divided into subclasses according to their outburst duration.  Long duration transients, which are visible for weeks or even months, are either associated with HMXBs (high-mass X-ray binaries, containing a Be star as the donor, see Charles \& Coe 2006) which display hard X-ray spectra,  or LMXBs (low mass X-ray binaries, containing a low-mass, late-type K-M star as donor) and with soft X-ray spectra (Tanaka 2008).  There are also very short duration X-ray transients which are only visible for hours to a few days, which are believed to be associated with supergiant HMXBs (Sguera et al 2007).

Our understanding of these systems and the general properties of irradiated accretion discs has been greatly enhanced by the improved quality of long-term X-ray monitoring data that are now available.  Online archive datasets, especially that provided by RXTE, which now span more than 10 years, allow for detailed investigation of long-term variations in the lightcurves of X-ray sources.

Long-term, non-orbital periodicities, ranging from 10 to 100 days, were first discovered in the early days of X-ray astronomy (see Charles et al 2008 for a recent review).  Some of these are remarkably stable (e.g. those of Her X-1 and LMC X-4), whereas others (e.g. SMC X-1 and Cyg X-2) show clearly quasi-periodic variations over the range of 50-200 days. These superorbital periods are thought to be related to the properties of the accretion disc, which include radiation-induced warping (Ogilvie \& Dubus 2001) and/or precession (Whitehurst \& King 1991) of the accretion disc. Radiation pressure from the intense X-rays arising near the compact object causes the warping of the accretion disc, while tidal forces in high mass-ratio binaries lead to the precession of the accretion disc.  Both effects can lead to the periodic obscuration of the compact object, (Clarkson et al 2003a, Clarkson et al 2003b). 

However, there is a small group of X-ray sources that initially exhibit transient-like behaviour in their X-ray lightcurves, but then subsequently remain in an ``on'' or active state and do not return to their pre-outburst level.  These ``pseudo-transients'' can go on to display further intervals of enhanced activity (e.g. McClintock \& Remillard 2006, and references therein). 

There is no single classification scheme for transients and their labelling tends to be a function of when they were discovered. Some sources were classified transients, while they may actually be pseudo-transients if long-term variability in their lightcurves is taken into consideration.  Their classification as transients is simply due to the relatively short period of time over which extended X-ray observations have been undertaken.

Here we report on an analysis of the 10-year X-ray history of a new transient source, IGR J17098-3628 \footnote{This source is designated IGR J17098-3626 on the RXTE/ASM website, but is referenced as such only once in the literature (Kennea et al 2005) when in fact referring to IGR J17098-3628.}. This source was discovered with INTEGRAL (Grebenev et al 2005) as part of its regular monitoring of the Galactic Plane and Galactic Centre, searching for new and variable X-ray sources. 

IGR J17098-3628 lies too close to IGR J17091-3624 (another new INTEGRAL transient source that is $\sim$ 10 arcmins away) for RXTE ASM to resolve them (Chen et al 2008), but they are clearly resolved by INTEGRAL (Grebenev et al 2005, Grebenev et al 2007). The ASM light curve for IGR J17098-3628 therefore would represent the combined flux of both INTEGRAL sources. Both these sources are in close proximity ($\sim$ 1$^{\circ}$ away) to GX349+2, a bright but steady LMXB a.k.a. Sco X-2 (Grebenev et al 2007), which may affect our ASM light curve (Chen et al 2008). We will show, however, that in spite of these problems, our variability analysis has been able effectively to separate these components.

Initially, IGR J17098-3628 appeared to be a ``classic'' X-ray transient, but the ASM monitoring has shown that it remains (to the time of writing) in an active or ``on'' state.  We investigate its variability properties and compare them with those obtained for EXO 0748-676, the prototype for this behaviour (Parmar et al 1986).

\section{Observations and Data Analysis}

\subsection{RXTE/ASM}

The All Sky Monitor (ASM) aboard the Rossi X-ray Timing Explorer Project (RXTE) is provided by the Massachusetts Institute of Technology (MIT). The ASM has observed the X-ray sky since early 1996.  The ASM scans approximately 80\% of the sky during each orbit of the satellite, providing monitoring of a source every $\sim$90 minutes for at least 90 seconds. 
Each of the individual monitoring sessions is referred to as a ``dwell''. The ASM contains three rotating Scanning Shadow Cameras (SSC), allowing positional measurement of previously unknown sources to within 3 arcmin precision.

Data are available for source intensities in four energy bands: 1.5-3 keV (A), 3-5 keV (B), 5-12 keV (C) and a Sum band 1.5-12 keV (A+B+C). Data for each energy band are reduced independently by taking background effects into account and made available as dwell-by-dwell or one-day-averages, which are average counts constructed for each day from dwell-by-dwell data. Data are reduced and compiled weekly by the ASM team and made publically available on the ASM website: http://xte.mit.edu/ASMlc.html.  For full details see Levine et al (1996).

The ASM monitoring began on 1996 February 20, and at the time of this paper we were able to employ datasets spanning more than 10 years.  These contain the long-term intensity history of all X-ray sources in the RXTE catalogue, and this can be particularly valuable for examining the behaviour of an X-ray transient \textit{prior} to its initial discovery, as frequently weaker activity in the transient could easily have been missed. The continuous monitoring provided by ASM is therefore essential for appropriately classifying ``steady'' and ``transient'' behaviour.

\subsection{Data Reduction}

\begin{figure*}
  \centering
	\includegraphics[angle=0,width=0.78\textwidth]{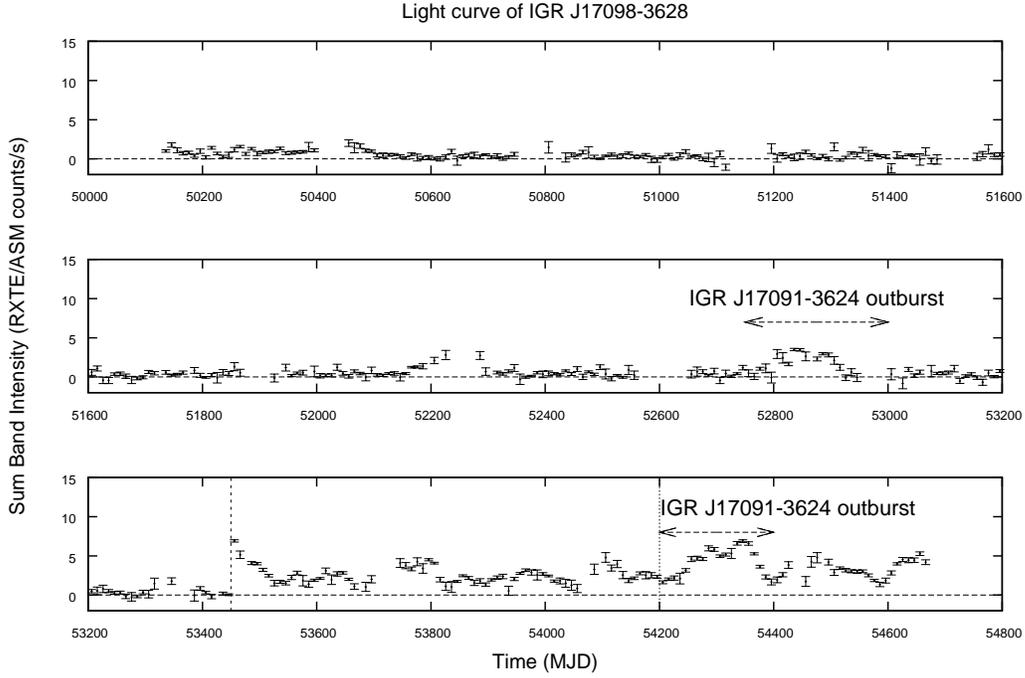}
  \caption {One-day average RXTE/ASM sum band lightcurve of IGR J17098-3628 (using 10 day data binning). The horizontal dashed lines indicate the zero flux level and are offset from the time-axis for clarity. The vertical dashed lines indicate the approximate start and end of the ``active'' state for IGR J17098-3628, which spans MJD 53450-54200. The ``quiescent'' state is prior to MJD 52500. Note the clear additional events that occurred during the intervals MJD 52750-53000 and MJD $>$ 54200 which are due to outbursts of the separate source IGR J17091-3624, as is discussed in the text.}
  \label{17098LightCurve}
\end{figure*}

\begin{figure*}
  \centering
	\includegraphics[angle=0,width=0.78\textwidth]{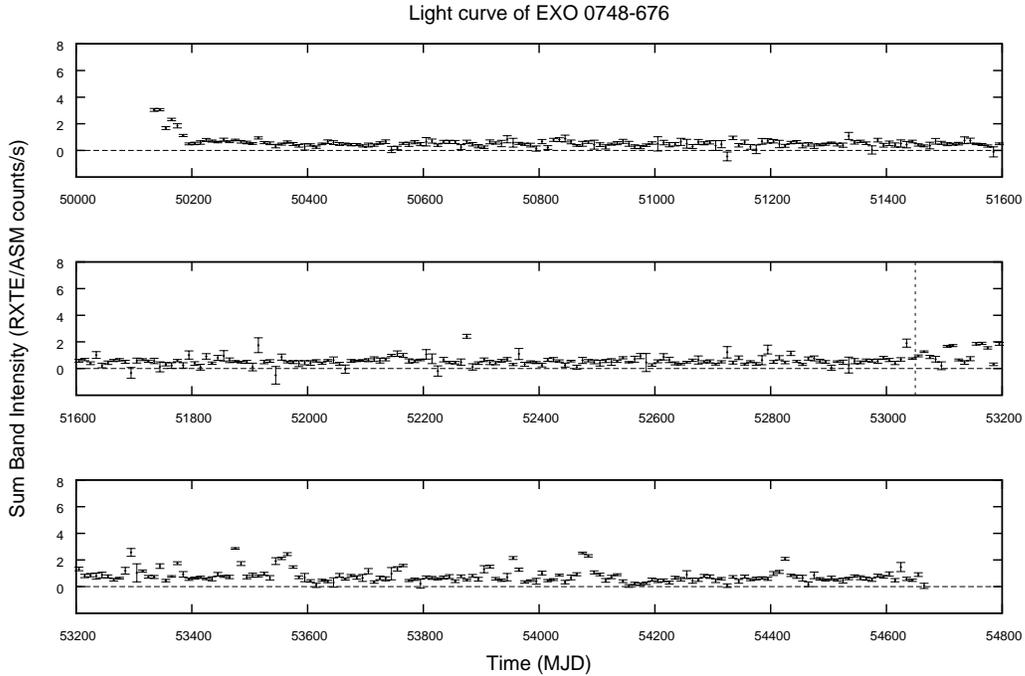}
  \caption {One-day average RXTE/ASM sum band lightcurve of EXO 0748-676 (using 10 day data binning). The horizontal dashed lines indicate the zero flux level and are offset from the time-axis for clarity. The vertical dashed lines indicate the approximate start of the ``active'' state for EXO 0748-676 at MJD 53050. The ``quiescent'' state is prior to MJD 53050 and excludes the event at the beginning of the RXTE mission (i.e. prior to MJD 50200).} 
  \label{0748LightCurve}
\end{figure*}

The full ASM lightcurves of IGR J17098-3628 and EXO 0748-676 from 1996 (MJD 50136) to the present (MJD 54678, i.e. JD-2400000.5), are displayed in Figures \ref{17098LightCurve} and \ref{0748LightCurve} respectively. Both these lightcurves show significant structure.

Note that EXO 0748-676 is still detected in its ``inactive'' or ``steady'' state where flux values are not zero, but do remain steady. Though it is not a ``quiescent'' state by the strict definition of the term, we will refer to it as the ``quiescent'' state of EXO 0748-676 in this paper. The ``active state'' contains additional flux that is superposed on the ``quiescent'' state flux level. An apparent outburst of EXO 0748-676 very early in the RXTE mission (prior to MJD 50200), will be excluded from our analysis of its ``quiescent'' state, which otherwise includes all data before the ``active'' state (which begins at MJD 53050).

An apparent precursor to the ``active'' state for IGR J17098-3628 occurs at MJD 52750-53000 and will be excluded from our analysis of the ``quiescent'' state. However, this ``precursor'' is in fact due to the outburst of the unrelated source IGR J17091-3624, which as already mentioned, INTEGRAL could clearly resolve from IGR J17098-3628 (Grebenev et al 2007). IGR J17091-3624 was again in outburst from MJD 54200 (Capitanio et al 2009), so data from MJD $>$ 54200 have also been excluded from the IGR J17098-3628 ``active'' state. Furthermore, Capitanio et al determined from subsequent monitoring by XMM, Swift and INTEGRAL, that only IGR J17098-3628 was active for MJD 53450-54200 and that IGR J17091-3624 therefore appears to have been quiescent between its outbursts, leaving us with $>$ 700 days of data on IGR J17098-3628 alone during its ``active'' state.

\subsection{Variability Analysis}

As discussed above, the one-day average ASM datasets were divided into ``active'' and ``quiescent'' states. Fourier analysis was employed to search for periodicities in the linearly detrended one-day average ASM sum-band lightcurve data. 

Since the Lomb-Scargle method (L-S) is useful for detecting weak periodic signals in unevenly sampled time-series data (Scargle 1982), we used it in our variability analysis of the ASM lightcurve data. For our analysis we made use of the PERIOD software within the Starlink Software Collection (http://www.starlink.rl.ac.uk). 

\begin{figure}
  \centering
	\includegraphics[angle=0,width=0.47\textwidth]{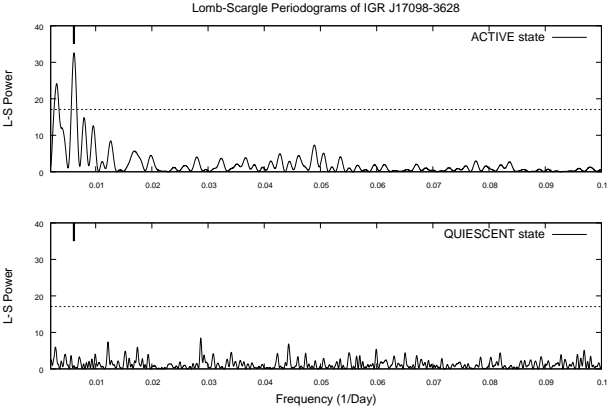}
  \caption {Lomb-Scargle periodograms of one-day averaged RXTE/ASM sum band lightcurve of IGR J17098-3628.} 
  \label{17098LS}
\end{figure}

\begin{figure}
  \centering
	\includegraphics[angle=0,width=0.47\textwidth]{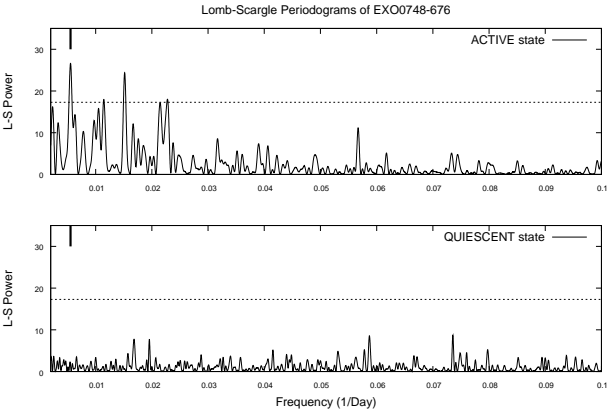}
  \caption {Lomb-Scargle periodograms of one-day averaged RXTE/ASM sum band lightcurve of EXO 0748-676.}
  \label{0748LS}
\end{figure}

The period range searched was 2-1000 days for the ``quiescent'' states and 2-500 days for the ``active'' states.
L-S periodograms of IGR J17098-3628 and EXO 0748-676 are displayed in Figures \ref{17098LS} and \ref{0748LS} respectively for both states, and span the frequency range 0.002-0.1 day$^{-1}$. (i.e. periods of 10-500 days).

The value of the L-S power above which we consider the signal in the periodogram statistically significant, is based on the results of a significance test. The latter involves a montecarlo simulation, whereby $10^{4}$ random datasets are generated while assuming white noise distribution and using the same time values, mean and standard deviation as contained in the lightcurve data. Periodograms according to the L-S technique are then calculated for each of these random datasets in order to determine the period with the highest power. The distribution of these powers results in a cumulative probability distribution function, whereby the power associated with a particular probability can be determined.

For each source, the respective 99.99\% confidence level is indicated by a horizontal line on each of their L-S periodograms. A power on the L-S periodogram above this confidence level indicates that the signal has a significance of $>$99.99 \% and can therefore be considered a potentially significant period in the lightcurve data.

IGR J17098-3628 and EXO 0748-676 display no significant periodicity in the data taken during their respective ``quiescent'' states.  However, significant periodic signals are present during their ``active'' states and include dominant periods of 162.9 $\pm$ 0.3 days for IGR J17098-3628 and 180.8 $\pm$ 0.3 days for EXO 0748-676, to a 99.99 \% degree of certainty.
The Phase Dispersion Minimisation (PDM) method also yielded the same results. 
It is important to also note that we see no such periodic signatures in these sources' extensive quiescent intervals.
There are 4.6 superperiod cycles covered in the active state of IGR J17098-3628, while 9 superperiod cycles are covered in the active state of EXO 0748-676.

Furthermore, in an effort to eliminate a possible systematic effect, identical analysis of known steady X-ray sources (e.g. the Ophiuchi galaxy cluster and the Vela pulsar), revealed no long-term signals in excess of the noise level. The Ophiuchi galaxy cluster is an extended source of comparable, non-variant luminosity in approximately the same sky region as IGR J17098-3628. The Vela pulsar is a source of comparable, non-variant luminosity on timescales of $>$1d and is located in approximately the same sky region as EXO 0748-676.

Identical analysis of GX349+2 revealed no long term periodic variations in excess of the noise level. More importantly, the absence of any signals during the ``quiescent'' state for IGR J17098-3628, rules out any contribution from GX349+2. Therefore the periodicity in the ASM light curve for IGR J17098-3628 in the ``active'' state, is not the result of any variability of GX349+2. As discussed in section 2.2, we believe that our ASM data for the ``active'' state of IGR J17098-3628 are uncontaminated by IGR J17091-3624, as INTEGRAL observations showed that the latter was quiescent. The periodic signal contained in our ASM light curve during this time interval, we therefore attribute entirely to IGR J17098-3628.

\subsubsection{Phase-binned lightcurves}

\begin{figure}
  \centering
	\includegraphics[width=0.41\textwidth]{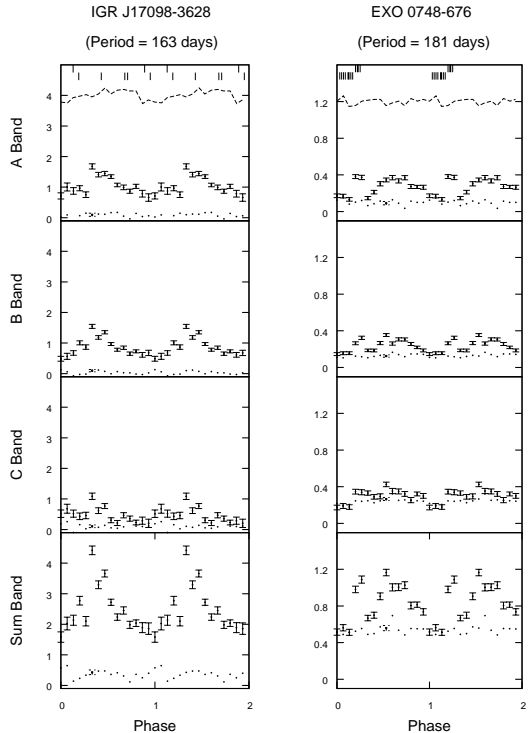}\\
  \caption {Phase-folded lightcurves of IGR J17098-3628 (left) and EXO 0748-676 (right) for their ``active''(with errorbars) and ``quiescent''(points) states. For the latter, the average error in each energy band is plotted as an errorbar on one point for reference. See text for explanation of the tick marks and dashed curve in the top box.} 
  \label{17098n0748Phase}
\end{figure}

Phase-folded lightcurves (Figure \ref{17098n0748Phase}) were constructed for all the energy bands by folding the data (using 15 phase bins) on the dominant significant period determined from the Sum band data for each source. The ``active'' and ``quiescent'' states are plotted together and the ``quiescent'' data are very different for the sources. In order to avoid confusion, the ``quiescent'' states are plotted without errorbars, but for clarity the average errors for each band were plotted on one point in each figure. The flux for EXO 0748-676 in the ``quiescent'' state is fairly steady, but is above zero as discussed in section 2.2. The flux for IGR J17098-3628 in the ``quiescent'' state contains a weak signal due to systematic effects, probably introduced by its proximity to GX349+2, but the level of detection is marginal. Therefore, the flux for IGR J17098-3628 in the ``active'' state lies significantly ($\sim$ 10 times brighter) above its ``quiescent'' state flux, while the flux for EXO 0748-676 in the ``active'' state shows a non-sinusoidal periodic variation superposed on the steady, non-zero flux level of its ``quiescent'' state. The latter effect is apparent in the A and B bands, but not the C band for EXO 0748-676. 

The phase-folded lightcurves clearly show the periodic behaviour in our ASM data for the ``active'' states of IGR J17098-3628 and EXO 0748-676. The $\approx$ 66 d period present in our ASM data for EXO 0748-676 introduces clear additional structure in the phase-folded lightcurves.

Two further tests were included to check the validity of the periodic signals, since it is possible that periods close to $\approx$ 0.5 yr or 1 yr might be related to the angular proximity of the sources to the Sun. When the angular separation between the Sun and the source is $<$30$^{\circ}$, annual gaps may occur in the data and since the Sun is bright in X-rays, source flux levels may also be elevated by scattered solar X-rays entering the detector. Correlation between maxima in the phase-folded lightcurves and the occurrence of minimum Solar angular separation, would therefore indicate a possible false period introduced by the angular proximity of the Sun. The minimum Solar angular separation angles are $\sim$14$^{\circ}$ for IGR J17098-3628 and $\sim$80$^{\circ}$ for EXO 0748-676, so that the effect of the Sun would be relevant to the former but not the latter. As expected, annual sampling gaps for IGR J17098-3628 occur around the times of minimum Solar angular separation, while not occurring for EXO 0748-676 (which is continuously viewable by RXTE). The times of the annual minimum Solar angular separation were plotted as tick marks in the top box of the phase-folded lightcurves, where the first and second rows are associated with the active and quiescent states of the sources respectively. The minimum Solar angular separation during the active state (tick marks in the top row) are only of particular interest in IGR J17098-3628 and these are sufficiently scattered in phase and uncorrelated to the peaks in the phase-folded lightcurves (Figure \ref{17098n0748Phase}), to rule out the effects of the Sun as a possible cause for the observed periods.  Data sampling rates (normalised to fit on the figures) are represented by dashed lines at the top of Figure \ref{17098n0748Phase}. Since there appears to be no correlation between sampling dips and the peaks in the phase-folded lightcurves, we can rule out non-uniform data sampling as a possible cause for the observed periods. 

A larger contribution of flux in the lower energy bands in both sources during the ``active'' state, is apparent in their phase-folded lightcurves. This effect is also apparent in the average flux values, summarised in Table \ref{averageflux}. For both sources, the increase in flux from the ``quiescent'' state to the ``active'' state, contains a larger contribution from the lower energy bands, implying that the sources became softer.

We also include in table \ref{averageflux} the $\chi_{\nu}^{2}$ values (in brackets) for a constant fit to the ``quiescent'' state phase folded lightcurves of Figure \ref{17098n0748Phase}. The sum band $\chi_{\nu}^{2}$ values for the two sources are comparable and provide evidence at the $99.9\%$ level for significant variability in these data. 

The excess flux events, occurring before MJD 50500 and at MJD 52150-52350, were excluded from the average flux calculation for IGR J17098-3628. These events are possibly contamination of the ASM lightcurve by IGR J17091-3624 and were subsequently also removed from the phase-folded lightcurves for IGR J17098-3628.

\begin{center}
	\begin{table}
	\caption{Average ASM Flux values of IGR J17098-3628 and EXO 0748-676}
	\label{averageflux}
	\smallskip
	\begin{tabular}{lcc}
	\hline
	\bf{IGR J17098-3628} \\
	Energy & ``active'' & ``quiescent'' \\
	band & state & state \\
	 &  &  ($\chi_{\nu}^{2}$ for constant flux)\\
	\hline
	Sum & $2.52 \pm 0.07$ & $0.36 \pm 0.04\ (3.3)$  \\
	A   & $1.06 \pm 0.04$ & $0.07 \pm 0.03\ (2.8)$ \\
	B   & $0.88 \pm 0.03$ & $0.03 \pm 0.01\ (2.7)$  \\
	C   & $0.47 \pm 0.04$ & $0.08 \pm 0.02\ (4.0)$  \\
	\hline
	\bf{EXO 0748-676}\\
	Energy & ``active'' & ``quiescent'' \\
	band & state & state \\
	 &  &  ($\chi_{\nu}^{2}$ for constant flux)\\
	\hline
	Sum & $0.83 \pm 0.02$ & $0.54 \pm 0.01\ (3.5)$  \\
	A   & $0.27 \pm 0.01$ & $0.10 \pm 0.01\ (2.3)$  \\
	B   & $0.24 \pm 0.01$ & $0.13 \pm 0.01\ (1.8)$  \\
	C   & $0.30 \pm 0.01$ & $0.24 \pm 0.01\ (2.9)$  \\
	\hline
	\end{tabular}
	\end{table}
\end{center}

\section{Discussion}

\begin{center}
	\begin{table}
	\caption{Comparison of IGR J17098-3628 and EXO 0748-676}
	\label{compare}
	\smallskip
	\begin{tabular}{lcc}
	\hline
	\bf{Source} & \bf{IGR J17098-3628} & \bf{EXO 0748-676}\\
	\hline
	$P_{orb}$ &  ? & 3.8 hrs $^{[1]}$\\
	V & $\sim$21 $^{[2]}$&  16.9--17.5 $^{[3]}$\\
	ASM count rate & 2--4 c/s & 0.5--1.2 c/s \\
	$N_X$ cm$^{-2}$ & 8.9$\times$10$^{21}$ $^{[4]}$ & 4.1$\times$10$^{20}$  \\
	E(B-V) & 1.3 & 0.06 $^{[5]}$\\
	$A_V$  & 3.9 & 0.18 \\
	Active state  & MJD 53450--54200 & MJD 53050--54677 \\
	$P_{sup}$ & 163 d & 181 d \\  
	$P_{sup}$ cycles  &  4.6  &  9.0 \\
	\hline
	\end{tabular}	
	\footnotesize{$^{[1]}$ Parmar et al 1986, $^{[2]}$ Grebenev et al 2007, $^{[3]}$ Schoembs et al 1990, $^{[4]}$ Kennea et al 2007, $^{[5]}$ Hynes et al 2006}
	\end{table}
\end{center}

The quantity and quality of the ASM archive has stimulated significant interest in the long-term variability properties of X-ray binaries.  This can be seen in the overview articles by Wen et al (2006), Sood et al (2007) and Charles et al (2008), with the last describing at least 6 different mechanisms by which superorbital periodicities (observable in X-ray and/or optical light-curves) can manifest themselves.  Unfortunately, we do not yet know the orbital period or even nature of the donor star in IGR J17098-3628, but the similarity of its long-term X-ray behaviour to EXO0748-676 suggests that we can infer what these properties might be.  
Table \ref{compare} compares these two sources and we estimate reddening values using standard relations 
($A_V = 3 E(B-V)$ and $N_X = 6.8 \times 10^{21} E(B-V)$ [cm$^{-2}$]). 

The suggested optical counterpart of IGR J17098-3628 (Steeghs et al 2005) is much fainter than that of EXO0748-676, which may appear surprising given their very similar observed X-ray count rates.  However, a key difference between the sources is their galactic latitude and hence levels of interstellar extinction, with IGR J17098-3628 suffering a much higher intrinsic column density (Kennea et al 2007).  This can be translated into an E(B-V) of 1.3, and hence $A_V$ that is almost 4 magnitudes, accounting nicely for the observed differences in optical brightness.  Furthermore, this implies that the companion star and/or disc in IGR J17098-3628 is similar to that in EXO0748-676 and hence is also an LMXB in a relatively short period ($<$1d) binary system.

It is then of interest to note that EXO0748-676 also demonstrates the presence of a superorbital modulation when in its active state.  If both systems are LMXBs with at least neutron star compact objects (that is certainly the case for EXO0748-676 as it is an X-ray burster, see e.g. Hynes et al 2006) then the mass ratios ($M_X/M_2$) will be high ($\geq$3).  This means that they will be susceptible to precession of the accretion disc as is seen in similar mass ratio cataclysmic variables (Whitehurst \& King 1991).  But the LMXBs are also X-ray luminous when in their active state, and so X-ray irradiation of the disc can lead to sustained warping. When precession and warping are coupled, the resultant precessing warped accretion disc would manifest itself as the observed superorbital periods. These are types 2, 3 and 5 in the categorisation of Charles et al (2008).

However, with no detected pulsations or bursts, the nature of the compact object in IGR J17098-3628 remains undefined. We note though, that its intrinsic X-ray spectrum (corrected for absorption) is softer than that of EXO0748-676 and so it may well be a black hole (BH) system.

\section{Acknowledgements}
ASM results were provided by the ASM/RXTE teams at MIT and at the RXTE Science Operations Facility and Guest Observer Facility at NASA's Goddard Space Flight Center (GSFC). The ASM results were provided by the ASM/RXTE teams at MIT and at the RXTE Science Operations Facility and Guest Observer Facility at NASA's Goddard Space Flight Center. We would like to thank Dr. Alan Levine for his comments and suggestions on a previous version of this manuscript and also the use of his fortran program for the significance test. We also thank the anonymous referee for comments that have helped clarify our presentation of the effects of sampling gaps in the data.

\label{lastpage}

\end{document}